\documentclass[a4paper]{appolb}
\usepackage{graphicx}
\usepackage{hyperref}
\usepackage{amsmath}
\usepackage[numbers]{natbib}
\usepackage{booktabs}

\newcommand{\pp}{\ensuremath{\text{p\kern-0.05em p}}}

\newcommand{\PbPb}{\ensuremath{\mbox{Pb--Pb}}}
\newcommand{\AuAu}{\ensuremath{\mbox{Au--Au}}}
\newcommand{\sqrts}{\ensuremath{\sqrt{s_{\text{NN}}}}}

\newcommand{\figRef}[1]{Fig.~\ref{#1}}
\newcommand{\figureRef}[1]{Figure~\ref{#1}}

\newcommand{\GeVc}{\ensuremath{\text{GeV}\kern-0.05em/\kern-0.02em c}}

\newcommand{\pT}{\ensuremath{p_{\text{T}}}}

\newcommand{\Raa}{\ensuremath{R_{\text{AA}}}}

\newcommand{\qhat}{\ensuremath{\widehat{q}}}

\begin{document}
\title{Bayesian analysis of QGP jet transport using multi-scale modeling applied to inclusive hadron and reconstructed jet data%
\thanks{Presented at the 29th International Conference on Ultrarelativsitic Nucleus-Nucleus Collisions}%
}
\author{Raymond Ehlers${}^{1}$ on behalf of the JETSCAPE Collaboration
\address{${}^{1}$UC Berkeley, Lawrence Berkeley National Laboratory}
}
\maketitle
\begin{abstract}
The JETSCAPE Collaboration reports a new determination of jet transport coefficients in the
Quark-Gluon Plasma, using both reconstructed jet and hadron data measured at RHIC and the
LHC.
The JETSCAPE framework incorporates detailed modeling of the dynamical evolution of
the QGP; a multi-stage theoretical approach to in-medium jet evolution and medium response;
and Bayesian inference for quantitative comparison of model calculations and data.
The multi-stage framework incorporates multiple models to cover a broad range in scale of the in-medium
parton shower evolution, with dynamical choice of model that depends on the current virtuality
or energy of the parton.

We will discuss the physics of the multi-stage modeling, and then present a new Bayesian
analysis incorporating it.
This analysis extends the recently published JETSCAPE determination
of the jet transport parameter $\qhat{}$ that was based solely on inclusive hadron suppression
data, by incorporating reconstructed jet measurements of quenching.
We explore the functional dependence of jet transport coefficients on QGP temperature and jet energy and
virtuality, and report the consistency and tensions found for current jet quenching modeling with
hadron and reconstructed jet data over a wide range in kinematics and $\sqrts{}$.
This analysis represents the next step in the program of comprehensive analysis of jet quenching
phenomenology and its constraint of properties of the QGP.

\end{abstract}

\vspace{-0.5cm}

\hypertarget{introduction}{%
\section{Introduction}\label{introduction}}

The multitude of unfolded experimental jet quenching measurements from
RHIC and the LHC contains a wealth of information. Given the ability of
models to successfully describe individual measurements despite
utilizing different formulations of the underlying physical phenomena,
broader data-model comparisons are required to assess the ability of a
model to describe the full set of measured data. Is it currently
possible to make a consistent picture of all of these measurements? And
if we can construct such a picture, what physics can we extract?

To address these questions, we reframe our perspective to ask: for a
given model, what parameters are most compatible with experimental
measurements? When reframed in this inherently Bayesian manner, Bayesian
inference provides a clear approach to extract model parameters while
incorporating knowledge about both theory and experiment, including
their respective uncertainties.

Bayesian inference is performed by taking advantage of Bayes' theorem,

\begin{equation}\label{eq:bayesTheorem}
P(\theta | x) = \frac{P(x | \theta) P(\theta)}{P(x)},
\end{equation}

\noindent{}where \(x\) represents the data and \(\theta\) the model
parameters. This expression relates the prior distribution,
\(P(\theta)\), to the posterior distribution, \(P(\theta | x)\), via the
likelihood that the data is described by the model parameters,
\(P(x | \theta)\). In heavy-ion physics, Bayesian inference has been
successfully used in the soft sector
\cite{Bernhard:2016tnd,Novak:2013bqa,JETSCAPE:2020mzn,JETSCAPE:2020shq},
but its application to the hard sector -- as discussed here -- is less
well developed.

In these proceedings, we report the status of Bayesian inference with
jet quenching data using JETSCAPE. In order to explore our model using
Bayesian inference, the jet transport coefficient \(\qhat{}\) is
parametrized with a physics-inspired expression with externally tunable
parameters, which form an N-dimensional parameter space. Simulations for
any given set of parameters are computationally expensive since they
should ideally have uncertainties much smaller than the experimental
observables used for constraining the parameters. Consequently, full
exploration of the parameter space with the available computational
resources requires a careful strategy to maximize the combination of
statistical precision and coverage of the phase-space with precise
calculations. To address this issue, simulations are performed for a
representative set of ``design points'', and non-parametric
interpolation is performed via Gaussian Process Emulation. The parameter
space is then explored via Markov Chain Monte Carlo methods, utilizing
Bayes' theorem to determine the most probable model parameters.

\vspace{-0.3cm}

\hypertarget{bayesian-inference-with-inclusive-charged-hadron-raa}{%
\section{\texorpdfstring{Bayesian inference with inclusive charged
hadron
\(\Raa{}\)}{Bayesian inference with inclusive charged hadron \textbackslash Raa\{\}}}\label{bayesian-inference-with-inclusive-charged-hadron-raa}}

As a proof-of-principle of Bayesian inference in the hard sector, we
extracted model parameters using inclusive charged hadron \(\Raa{}\)
measurements at \(\sqrts{} =\) 200 GeV at RHIC and \(\sqrts{} =\) 2.76
and 5.02 TeV at the LHC as a function of centrality
\cite{JETSCAPE:2021ehl}. This analysis utilizes the JETSCAPE framework.
Jet propagation in the QGP was calculated using three different
formulations: MATTER, LBT, and an early version of the multi-stage
MATTER+LBT approach
\cite{Chen:2017zte,Cao:2017hhk,Majumder:2013re,Luo:2018pto,JETSCAPE:2017eso,Cao:2017qpx}.
In this analysis, the \(\qhat{}\) parametrization used in each model has
four or five parameters. The partons were propagated through
pre-computed 2+1D hydrodynamics events. Full details of the analysis are
described in \cite{JETSCAPE:2021ehl}.

\begin{figure}[t]
    \centering
    \includegraphics[width=0.7\textwidth]{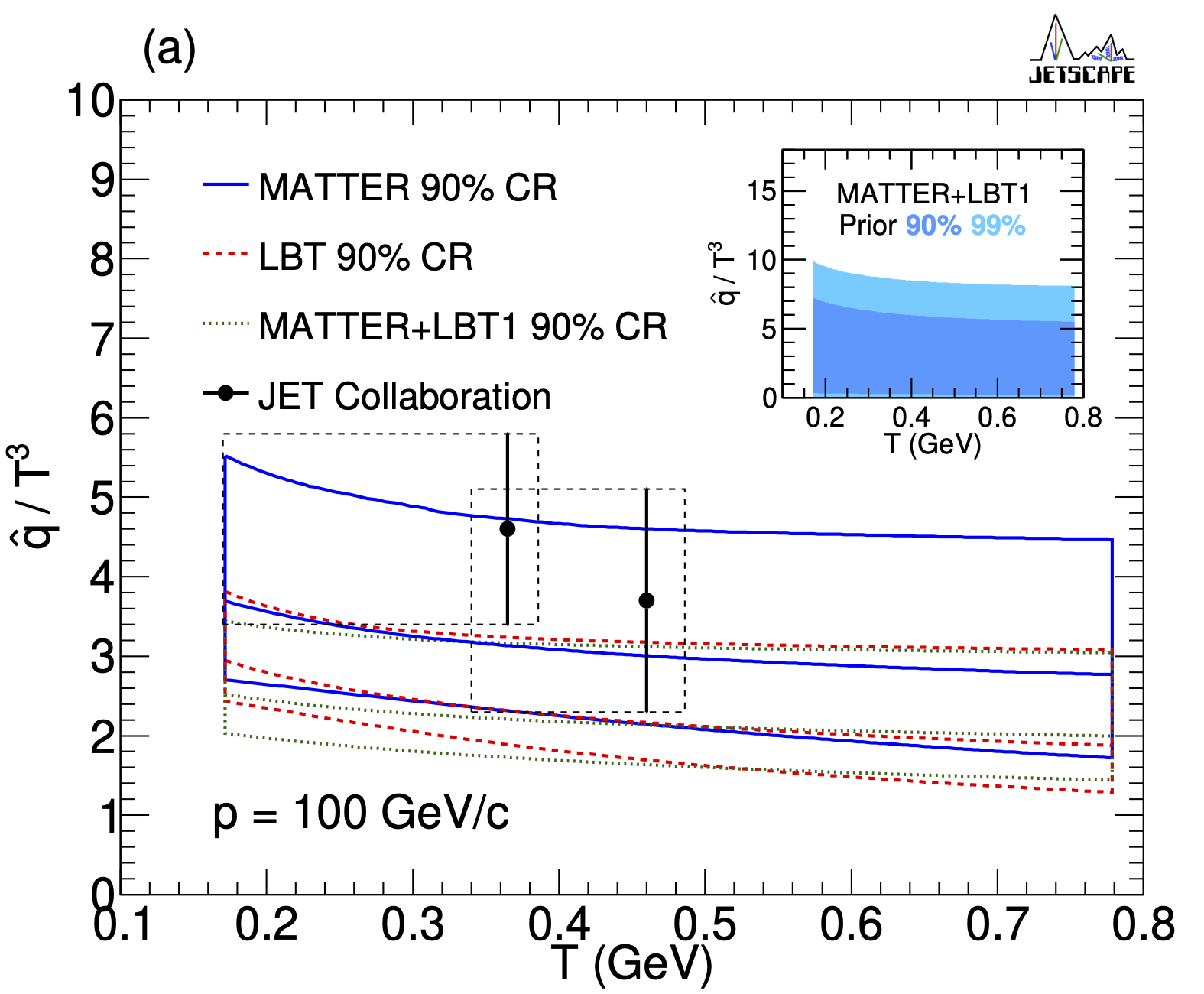}
    \caption{Posterior distribution of the temperatue dependence of the jet transport coefficient $\qhat{}$ extracted from inclusive charged hadron $\Raa{}$ measured at $\sqrts{}$ = 200 GeV, 2.76 TeV, and 5.02 TeV. The posterior is significantly constrained compared to the prior distribution.}
    \label{fig:qhatTemperatureDependenceHadronRaa}
\end{figure}

\figureRef{fig:qhatTemperatureDependenceHadronRaa} shows the posterior
distribution for the temperature dependence of \(\qhat{}\) from this
analysis. The extracted posterior is significantly constrained compared
to the prior distribution shown in the inset panel. Further, the values
from the three different models are consistent with each other, as well
as with the previous determination of \(\qhat{}\) by the JET
collaboration \cite{JET:2013cls}. This result demonstrates the viability
of Bayesian inference in the hard sector, setting the stage for more
complex analyses.

\vspace{-0.3cm}

\hypertarget{inclusive-jet-and-hadron-raa}{%
\section{\texorpdfstring{Inclusive jet and hadron
\(\Raa{}\)}{Inclusive jet and hadron \textbackslash Raa\{\}}}\label{inclusive-jet-and-hadron-raa}}

As the next step after this proof-of-principle analysis, there are many
possible directions. We elect to build on the previous analysis and
adiabatically add observables alongside the inclusive charged hadron
\(\Raa{}\). For this next analysis, we only add inclusive jet
\(\Raa{}\). We do not select data but take an agnostic approach,
including all relevant experimental measurements. This leaves it to the
Bayesian inference analysis to determine the model compatibility, as
well as to highlight possible experimental tensions.

For this analysis, we utilize a new multi-stage approach with MATTER+LBT
in the JETSCAPE framework \cite{JETSCAPE:2022jer}. This multi-stage
model includes the standard hard thermal-loop
\(\widehat{q}_{\text{HTL}_{}}\), modulated by additional coherence
effects at high virtuality, which are manifest via fewer interactions at
high \(Q^{2}\). The overall jet transport coefficient is defined as
\(\qhat{} = \widehat{q}_{\text{HTL}} \cdot f(Q^{2}_{})\), where

\begin{equation*}
f(Q^{2}) = \frac{N ( \exp{(c_{3} (1 - x_{B})} )}{1 + c_{1}\ln{(Q^{2}/\Lambda_{\text{QCD}}^{2})} + c_{2}\ln^{2}{(Q^{2}/\Lambda_{\text{QCD}}^{2})}},
\end{equation*}

\noindent{}and \(c_{1}\), \(c_{2}\), and \(c_{3}\) are parameters to be
extracted via Bayesian inference, and \(N\) normalizes the expression to
ensure that \(\qhat{} = \widehat{q}_{\text{HTL}_{}}\) once the model
transitions to LBT. In addition to the \(c\) parameters, \(\qhat{}\)
also depends on \(\alpha_{\text{s}}\), \(\tau_{0}\), and model switching
parameter \(Q_{\text{switch}}\), for a total of six parameters. Due to
the modular nature of the JETSCAPE framework, alternative models can be
included in future Bayesian analyses.

\begin{table}[t]
    \centering
    \begin{tabular}{@{}lll@{}}
    \toprule
    Experiment & $\sqrts{}$ & Inclusive $\Raa{}$ observables \\ \midrule
    STAR       & 200        & jets $R$ = 0.2, 0.4            \\
    PHENIX     & 200        & $\pi_{0}$ $\Raa{}$             \\
    ALICE      & 2.76, 5.02 & jets $R$ = 0.2, 0.4            \\
    ATLAS      & 2.76, 5.02 & hadron, jets $R$ = 0.4         \\
    CMS        & 2.76, 5.02 & hadron, jets $R$ = 0.2-0.4     \\ \bottomrule
    \end{tabular}
    \vspace{0.1cm}
    \caption{Experimental data from RHIC and the LHC included in the work-in-progress analysis. Although the full
    analysis will include all available data, we first focused on a subset of the measurements in 0-10\%.}
    \label{tab:experimentalData}
\end{table}

Since there are multiple measurements from individual experiments, as
well as additional systematic uncertainties related to jet analyses
(such as shape uncertainties, which tend to be anti-correlated), the
correlation of uncertainties requires particular care. The uncertainty
treatment generally follows the procedure from our previous analysis
\cite{JETSCAPE:2021ehl}. Correlated uncertainties that are not specified
in detail in the experimental publication are treated with a 10\%
correlation length.

\begin{figure}[t]
    \centering
    \includegraphics[width=0.7\textwidth]{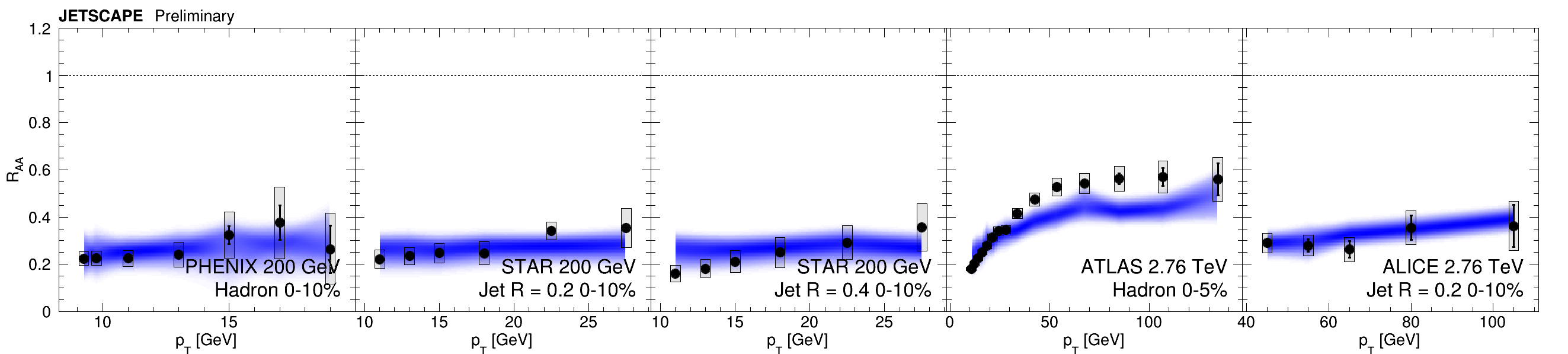}
    \includegraphics[width=0.7\textwidth]{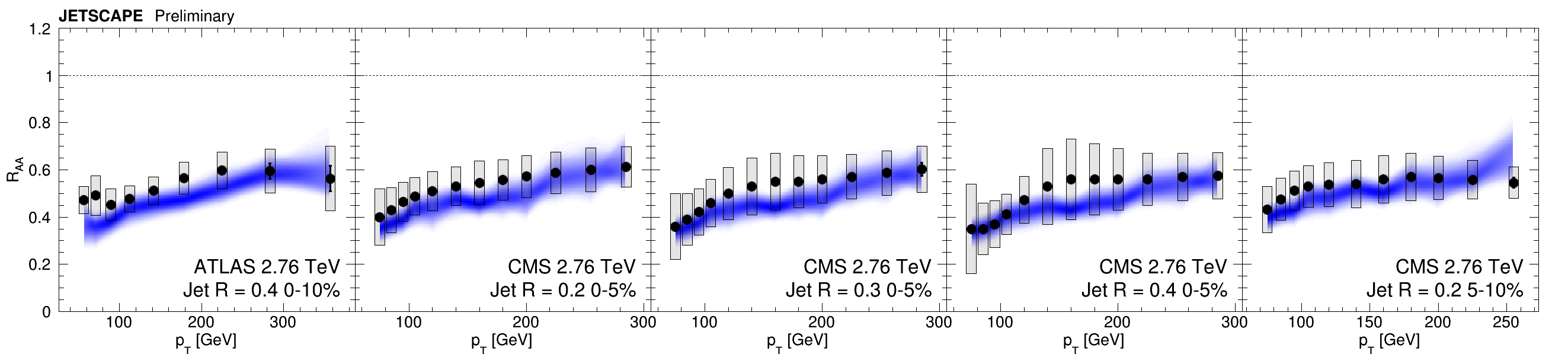}
    \includegraphics[width=0.7\textwidth]{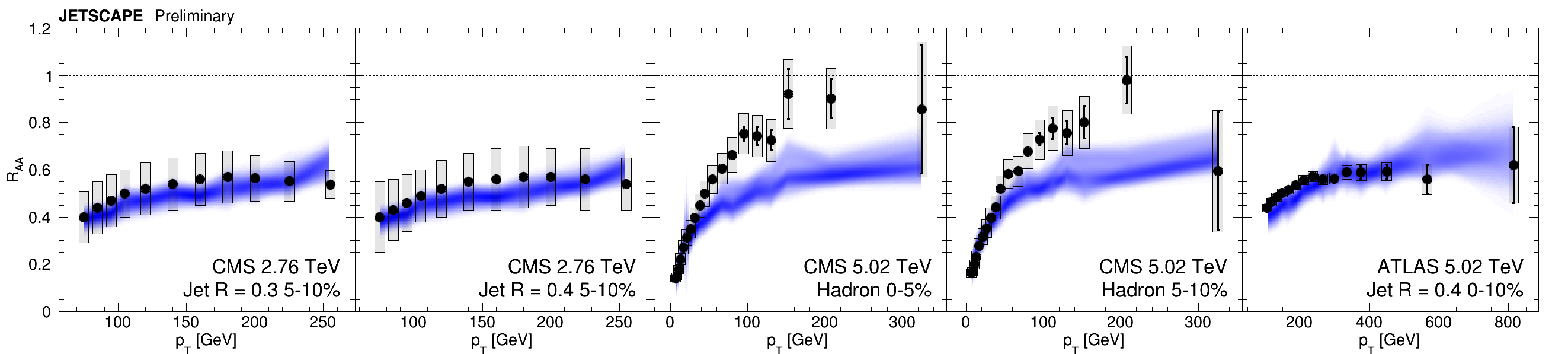}
    \includegraphics[width=0.7\textwidth]{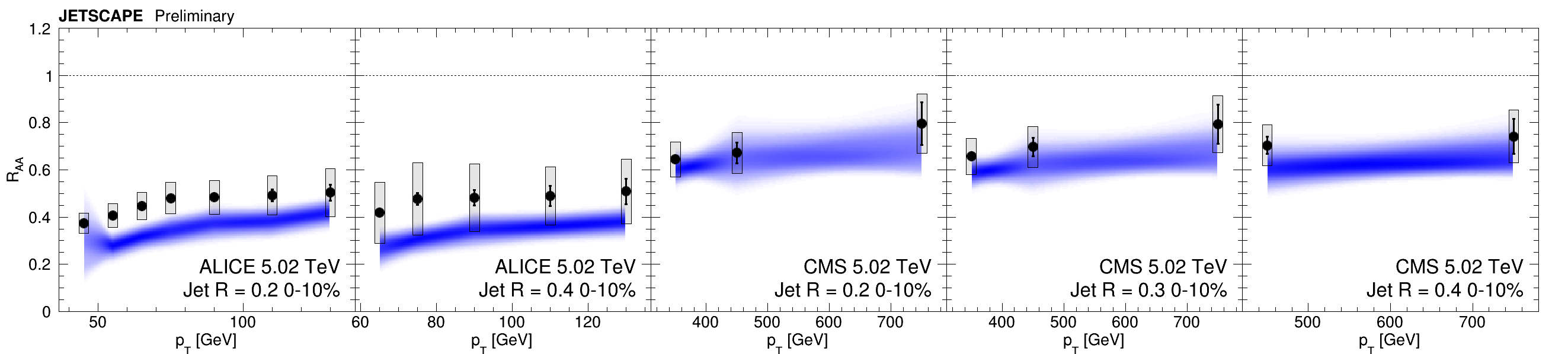}
    \caption{Posterior distribution of calibrated model parameters compared to inclusive jet and charged hadron $\Raa{}$
    data from RHIC and the LHC. The data are shown in black, while the sampled posterior is shown in blue. The model is able
    to describe the data fairly well overall, although there are some regions of tension.}
    \label{fig:hadronAndJetRaaPosterior}
\end{figure}

The work-in-progress results of this analysis are presented here, with
samples drawn from the posterior distribution shown in
\figRef{fig:hadronAndJetRaaPosterior}. The experimental data are shown
in black, with the bars for the statistical errors and the boxes for the
systematic uncertainties. The analysis shown here corresponds to only a
small subset of calculations currently in progress, which require
millions of CPU core-hours. High performance computing resources are
required in order to complete such calculations. We utilize XSEDE
computing resources \cite{xsede} for this result. For these proceedings,
we focus on 0-10\% most central \AuAu{} and \PbPb{} collisions,
calibrating the model against the data enumerated in Table
\ref{tab:experimentalData}. This is likewise only a subset of the
experiment data which will be used for the full analysis.

At a high level, the model is broadly consistent with the data. By
exploring the details of the posterior distribution shown in
\figRef{fig:hadronAndJetRaaPosterior}, a wealth of information can be
extracted. The model is consistent with the \(\sqrts{} =\) 200 GeV data
from PHENIX and STAR shown in the top row, although the constraining
power is somewhat limited due to the large uncertainties at higher
\(\pT{}\). At \(\sqrts{}=\) 2.76 and 5.02 TeV, the model is able to
describe the data over part of the \(\pT{}\) range, but there is
apparent tension between the model and the data at high \(\pT{}\).
Further tension is visible by focusing on the comparison of a single
hadron and jet \(\Raa{}\) measurement at fixed \(\sqrts{}\), with small
uncertainties in both measurements driving the posterior distribution in
different directions. This tension is most apparent at \(\sqrts{}=\)
5.02 TeV, where the posterior tends to describe the hadron \(\Raa{}\)
and underpredict the jet \(\Raa{}\) at lower \(\pT{}\), transitioning to
underpredict the hadron \(\Raa{}\) and describe the ATLAS jet \(\Raa{}\)
at higher \(\pT{}\). Comparing \(R=0.4\) jet \(\Raa{}\) at \(\sqrts{}=\)
5.02 TeV across ALICE, ATLAS, and CMS illustrate some potential tensions
between the measurements at high \(\pT{}\), although the posterior tends
towards ATLAS jets, correlating with the small uncertainties.

\vspace{-0.3cm}

\hypertarget{outlook}{%
\section{Outlook}\label{outlook}}

We presented two Bayesian inference analyses using jet quenching
measurements, utilizing the JETSCAPE framework. The analysis of
inclusive charged hadron \(\Raa{}\) provides new constraints on the
temperature and momentum dependence of \(\qhat{}\). We further presented
the first multi-messenger Bayesian analysis for jet quenching,
considering inclusive jet \(\Raa{}\) measurements with the hadron
\(\Raa{}\) data. The posterior distribution of this work-in-progress
analysis already demonstrates the wealth of information encoded in this
multi-messenger approach. The model broadly describes the data, with
some tension with the hadron and jet \(\Raa{}\).

The treatment of uncertainties is critical to constrain model
parameters, and we strongly encourage experimental collaborations to
report full covariance matrices for their uncertainties, or at minimum
to report the signs of individual uncertainties relative to the central
values. Once the current simulations are completed, we will fully
explore the parameter space and perform a Bayesian inference analysis to
determine the distribution of model parameters which best describe the
data. We will also explore the consistency, corroboration, and
discriminating power of observables and kinematic selections, as well as
expand to additional observables.

\vspace{-0.3cm}

\section*{Acknowledgments}

This work was supported in part by the National Science Foundation (NSF)
within the framework of the JETSCAPE collaboration, under grant number
OAC-2004571.

\vspace{-0.3cm}

\scriptsize
\bibliography{rehlers.QM2022.bib}

\end{document}